# Operation Scheme Optimizations to Achieve Ultra-high Endurance ($10^{10}$) in Flash Memory with Robust Reliabilities

Yang Feng, Zhaohui Sun, Chengcheng Wang, Xinyi Guo, Junyao Mei, Yueran Qi, Jing Liu, Junyu Zhang, Jixuan Wu, Xuepeng Zhan and Jiezhi Chen, *Senior Member, IEEE*

*Abstract*— Flash memory has been widely adopted as stand-alone memory and embedded memory due to its robust reliability. However, the limited endurance obstacles its further applications in storage class memory (SCM) and to proceed endurance-required computing-in-memory (CIM) tasks. In this work, the optimization strategies have been studied to tackle this concern. It is shown that by adopting the channel hot electrons injection (CHEI) and hot hole injection (HHI) to implement program/erase (PE) cycling together with a balanced memory window (MW) at the high-$V_{th}$ (HV) mode, impressively, the endurance can be greatly extended to $10^{10}$ PE cycles, which is a record-high value in flash memory. Moreover, by using the proposed electric-field-assisted relaxation (EAR) scheme, the degradation of flash cells can be well suppressed with better subthreshold swings (SS) and lower leakage currents (sub-10pA after $10^{10}$ PE cycles). Our results shed light on the optimization strategy of flash memory to serve as SCM and implement endurance-required CIM tasks.

## I. INTRODUCTION

As one of the main streams of nonvolatile memories (NVM), flash memory has been widely utilized as stand-alone memory with large capacity and embedded memory with high-speed operations. Though it has robust reliability and the capability to construct large memory arrays, endurance is still a key concern [1-2] that seriously obstacles its applications in storage-class memory (SCM) and computing-in-memory (CIM). Recently, high-performance NAND flash memory has been developed in commercial products, such as Z-NAND from Samsung and XL-Flash from Kioxia [3-4], which draws much more attention to SCM in the realm of memory-centric computing, primarily due to its ability to bridge the gap between DRAM and 3D NAND flash memory [5-6]. This rise in demand is driven by the ever-increasing need for faster and energy-efficient processing to deal with mass data [7-8]. Moreover, for memory-centric computing and CIM architectures, frequent weight modulating and endurance are critical factors in guiding feasible designs [9-10]. With high endurance and fast operation speed [11-12], it will be possible to utilize flash memory as SCM to bridge DRAM and high-performance NAND flash memories, and undertake endurance-required CIM tasks.

In this work, optimization strategies have been studied to improve the endurance of flash memory by focusing on the operation schemes. Channel hot electrons injection (CHEI) and hot hole injection (HHI) are used to perform program/erase (PE) for weight updating. Then, with the proposed high threshold-voltage ($V_{th}$) mode and electric-field-assisted relaxation (EAR) schemes, a record high endurance ($10^{10}$) is achieved with well-suppressed cell degradations, including the subthreshold swing (SS) and leakage currents, which is minimized to sub-10pA even after $10^{10}$ PE cycles.

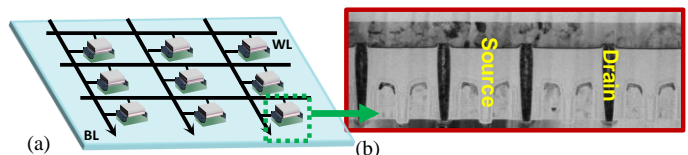

Fig. 1. (a) A schematic of the flash array. (b) TEM images of the flash array.

## II. OPTIMIZATION STRATEGIES FOR FLASH

In this work, characterizations are done in 55nm NOR flash memory. The tested memory array is shown in Fig.1, and the proposed optimization strategies for endurance are shown in Fig. 2. Firstly, as the weight updating scheme, CHEI and HHI are adopted to suppress the cell degradation based on previous work [13], which is also beneficial for fast and precise $V_{th}$ adjusting. Secondly, 20ns pulse width is used in each single pulse during PE cycling. Although a wider pulse can be used with a lower stress bias, it will sacrifice the speed of operation on the contrary. In addition, high-speed operation with a shorter but higher pulse might be helpful to suppress the degradation [14]. Moreover, in the conventional operation scheme towards storage usage, MW needs to be large to minimize the error bits. However, as for endurance-required applications, we can optimize MW to enhance the operation speed and lower cell degradation with fewer pulses and lower biases. In this work, the endurance exceeding $10^{10}$ cycles is demonstrated in flash cells by lowering the MW to 0.2V (the minimum tolerated MW). To understand the influence of CHEI and HHI, the statistics of SS and leakage current ($I_{off}$) are characterized by applying

This work was supported by National Natural Science Foundation of China (Nos. 62034006, 92264201, 62104134, 91964105), China Key Research and Development Program under Grant (2022YFB3603900), Natural Science Foundation of Shandong Province (ZR2020JQ28), and Program of Qilu Young Scholars of Shandong University.

Yang Feng, Zhaohui Sun, Chengcheng Wang, Xinyi Guo, Junyao Mei, Yueran Qi, Jixuan Wu, Xuepeng Zhan,and Jiezhi Chen are with School of Information Science and Engineering (ISE), Shandong University, Qingdao, China. 266237 (email: chen.jiezhi@sdu.edu.cn).

Jing Liu, is with Key Laboratory of Microelectronic Devices and Integrated Technology, Institute of Microelectronics of Chinese Academy of Sciences, Beijing, China.

Junyu Zhang are with Neumem Co., Ltd, Hefei, China.

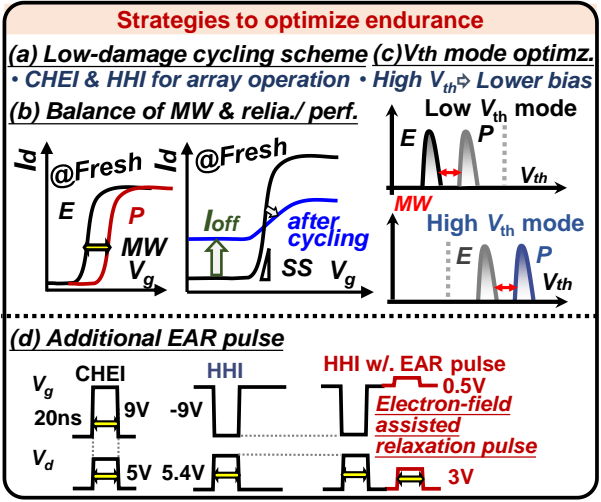

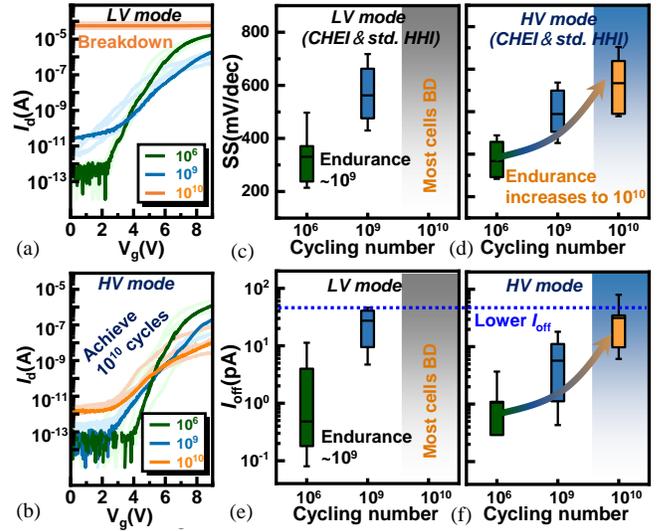

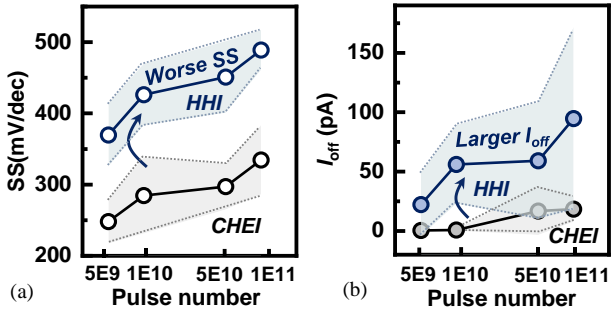

Fig. 2. (a) CHEI and HHI are adopted to lower the damage of weight updating; (b) the endurance can be extended by balancing the MW; (c) the difference between LV mode and HV mode, wherein the lower $V_{th}$ decreases programming voltage and increases erasing voltage; (d) the scheme of the electron-field assisted relaxation (EAR) pulse.

Fig. 4. IV curves in cells after different cycles at (a) LV mode and (b) HV mode. The statistical data at LV mode and HV mode, (c-d) SS, and (e-f) $I_{off}$. Cells are easier to break down after $10^9$ cycles at LV mode, while cells can achieve $10^{10}$ cycles at HV mode.

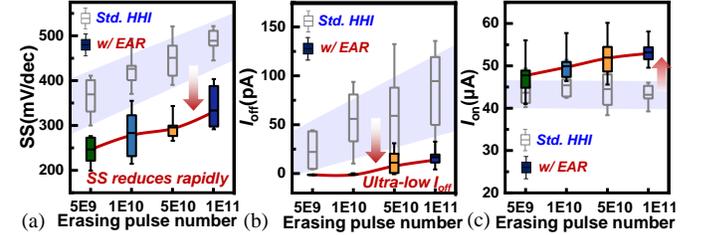

Fig. 3. The standard HHI pulses exhibit worse (a) SS and (b) $I_{off}$ compared with the CHEI pulses, lowering the voltage of HHI helps to enhance the endurance.

Fig. 5. The statistical data of (a) SS, (b) $I_{off}$, and (c) $I_{on}$ of multiple cells by repeating simple HHI with and without EAR. With the proposed EAR scheme, the degradations of cells can be suppressed with lower SS and lower $I_{off}$.

CHEI and HHI pulses to cells separately. As the statistical data shown in Fig.3, HHI stress causes larger degradation to cells obviously. This could be understood because the high-energy hole injection is more likely to generate traps in the oxide dielectrics [15]. Thus, lowering the damage from HHI stress could benefit endurance. In this consideration, the low $V_{th}$ (LV) mode and high $V_{th}$ (HV) mode are comparably studied with adjusted HHI biases. Here, we defined the LV mode in cells that are programmed to around 4V $V_{th}$, and the HV mode in cells that are programmed to around 7V $V_{th}$. Fig. 4(a-b) compares the IV curves in LV mode and HV mode, and it is observed that most flash cells are easier to break down after $10^9$ cycles in LV mode. For HV mode, the flash cells show obvious performance degradation but most cells can work with the increasing SS and higher $I_{off}$ before cycling. The performance of flash cells cycled with different modes is measured, and the statistics cycling results of the two modes are compared in Fig.4. Although the performance improvements of SS and $I_{off}$ are not large, the endurance of HV mode can easily achieve $10^{10}$ cycles. In cells at HV mode, more electrons are stored in the floating gate (FG). Thus the damage from HHI stress can be lowered because the erasing efficiency, including electron emission and electron-hole recombination, could be improved. Though the cells achieve $10^{10}$ cycles, the cells are still hard to apply to the SCM architectures as the result of the increasing SS and $I_{off}$.

Moreover, considering the possible damage by the charged holes in the oxide, a novel EAR scheme is designed to mitigate the cycling-related degradation, as described in Fig. 2. The additional EAR pulse is added after the HHI pulse to mitigate cell degradation. To evaluate the effect of the EAR pulse, the cells applied with and without the ERA pulse are tested. It is observed that, by adding the EAR pulse after the HHI pulse in HV mode, the degradation of SS and $I_{off}$ can be largely suppressed. Even after $10^{11}$ erasing pulses, the cells still exhibit acceptable SS (~334mV/dec), and ultra-low $I_{off}$ (~18.3pA), which are obviously better than the standard HHI pulse (Fig.5). For P/E cycled cells, adopting the proper CHEI pulse and HHI with EAR can effectively suppress the degradation of cells, as shown in Fig.6. The $I_d$-$V_g$ curves with the EAR pulse show better performance of SS, $I_{off}$ and $I_{on}$, which is more suitable to for SCM. The underlying physics of EAR effects can be explained by the degradation mechanisms of oxide dielectrics. During PE cycling, some high-energy holes are trapped in the tunneling oxide, which lowers the efficiency of further hole injection due to the higher tunneling barrier and can contribute to additional electron trap generation [16]. Therefore, according to the anode hole injection (AHI) model [17], hole injection is the dominant reason for the destructive breakdown in oxide. Thus, with the proposed EAR scheme, the trapped holes in the tunneling oxide could be released and the aforementioned damage can be suppressed to some extent. For the same reason, the MW of Fig. 6 can remain stable even after $10^{10}$ cycles. In a word, by implementing the optimized strategies, we can

## IV. CONCLUSIONS

This work demonstrated novel operation schemes to improve the endurance of flash memory. By adopting the developed optimization strategies combining the $V_{th}$ mode optimization and EAR scheme for SCM-target flash memory to mitigate the degradation of cells, a record high endurance ($10^{10}$) in flash is realized with effective suppression of SS and $I_{off}$ degradation, which can also make it feasible to utilize flash-based computing in memory architecture to process real-time high-precision computation tasks as the excellent weight updating abilities with the proposed scheme.

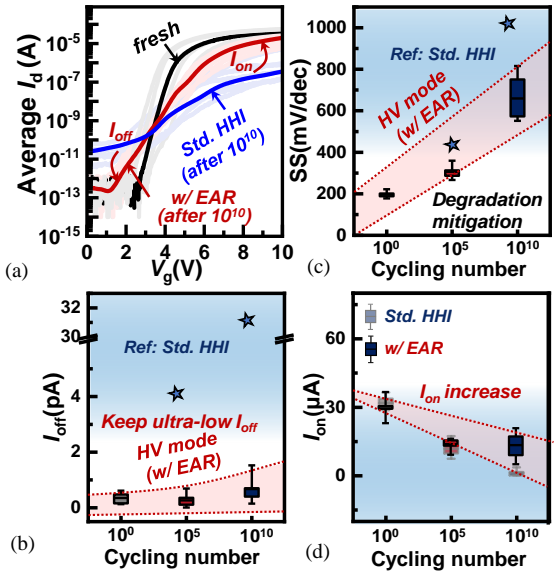

Fig. 6. (a) IV curves of fresh cells and cycled cells with different schemes. The statistical (b) $I_{off}$, (c) SS, and (d) $I_{on}$ in cells when cycling with CHEI and standard HHI, or HHI with EAR.

enhance the endurance to $10^{10}$ cycles while maintaining an acceptable level of degradation in SS, $I_{off}$, and $I_{on}$. In addition, this also lays the foundation for the realization of flash-based SCM applications that are capable of handling high-precision and large-scale data motion.

## III. RESULTS AND DISCUSSION

Besides the optimization strategies of endurance, stability is also an important factor in designing the robustness system and needs to be verified in the study. Considering the applied high voltage and the small MW, the memory state could probably change because the cells sharing the same bit line (BL) are programmed simultaneously. Here, drain disturb is tested in both HV mode and LV mode after adopting the EAR scheme. As shown in Fig.7(a), the HV mode shows smaller variations in µs-level, which is enough for ns-level HHI operations, manifesting that the HV mode also has good stabilities. Then, read disturb (RD) after $10^{10}$ P/E cycles are studied. For flash-based SCM systems, the short-term longevity ability is required to handle the I/O requests, thus the retention time requirements can be lowered accordingly [18-19]. In Fig.7(b), considering enlarged $V_{th}$ variations of cells after cycling, RD characteristic is tested in high-$V_{th}$ cells and low-$V_{th}$ cells. The high stability of read currents indicates that the proposed scheme can be well utilized in Flash-based CIM to ensure stable computing.

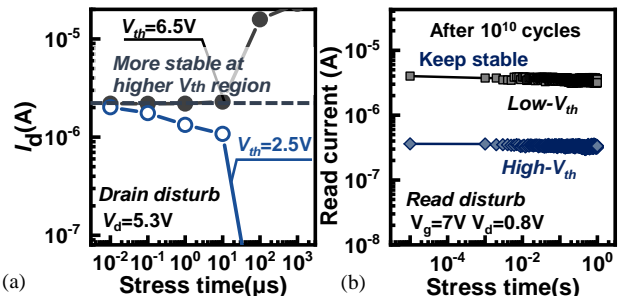

Fig. 7. (a) Current variations caused by drain disturb; (b) read current stabilities under constant read bias stressing.